\documentclass{PoS}
\title{Triple Higgs boson production at the ILC within a generic Two-Higgs-Doublet Model}
\ShortTitle{Triple Higgs boson production at the ILC within a generic Two-Higgs-Doublet Model}

\author{\speaker{Giancarlo Ferrera}%
\\
        Dipartimento di Fisica,
        Universit\`a di Firenze \& INFN Sezione di Firenze\\
        Via G. Sansone 1,
        I-50019 Sesto Fiorentino, Florence, Italy\\
        E-mail: \email{ferrera@fi.infn.it}}

\author{Jaume Guasch\\
        Gravitation and Cosmology Group,  Dept. FF \&  Institut de
        Ci\`encies del Cosmos, \\
        Universitat de Barcelona \\  Av. Diagonal 647, E-08028
    Barcelona, Catalonia, Spain\\
        E-mail: \email{jaume.guasch@ub.edu}}
\author{David L\'opez-Val\\
        High Energy Physics Group, Dept. ECM,\\
        Universitat de Barcelona\\
Av. Diagonal 647, E-08028 Barcelona, Catalonia, Spain\\
        E-mail: \email{dlopez@ecm.ub.es}}
\author{Joan Sol\`{a}\\
        High Energy Physics Group, Dept. ECM \&  Institut de Ci\`encies
        del Cosmos, \\
        Universitat de Barcelona\\
Av. Diagonal 647, E-08028 Barcelona, Catalonia, Spain\\
        E-mail: \email{sola@ifae.es}\vspace*{.4cm}}

\abstract{We present a study of triple Higgs boson (3H) production
at the International Linear Collider (ILC) within the general
Two-Higgs-Doublet Model (2HDM). We compute the production
cross-sections at the leading-order for the 3H final states and
find values up to $\sigma \sim 0.1\,$pb. This result represents a
large enhancement with respect to the corresponding MSSM
cross-sections, which stay typically at the level of $\sigma \sim
10^{-6}\,$ pb or less. Furthermore, since the 3H cross-sections in
the general 2HDM can be of the order of the double Higgs
production cross-sections, such 3H processes could be a competitive
(if not the dominant) mechanism for Higgs boson production at the
ILC. In practice, these 3H events could be identified through the
tagging of $6$ heavy-quark jet final states and, in this case,
they would provide strong evidence of an extended Higgs boson
sector -- likely of non-supersymmetric nature.\vspace*{.3cm}}

\FullConference{8th International Symposium on Radiative Corrections \\
                 October 1-5, 2007\\
                 Florence, Italy}

\begin{document}

 \section{Introduction}
 \label{Introduction}

Double Higgs boson (2H) production in a linear collider $e^+e^-
\to h^0\,A^0; H^0\,A^0; H^+H^-$, has been investigated in great
detail in the literature, although mainly in the Minimal
Supersymmetric Standard Model (MSSM)\,\cite{Djouadi:1992pu}.
However, a tree-level analysis of these pairwise-produced
unconventional Higgs bosons is most likely insufficient to
disclose their true nature. It is for this reason that a dedicated
work on radiative correction calculations on Higgs boson
production has been undertaken in various models
~\cite{mssmloop,ghk}. Besides, a second interesting mechanism for
studying the Higgs boson properties is the triple Higgs (3H)
production, which carries essential information to trace back the
ultimate structure of the corresponding Higgs potential. This kind
of processes have been profusely studied also within the MSSM,
although in this case the cross-sections are rather
meager\,\cite{tripleMSSM}. Our main purpose here is to study the
trilinear coupling $HHH$ in the general Two-Higgs-Doublet Model
(2HDM) by focusing on the 3H final states produced at the
ILC\,\cite{Ferrera:2007sp}:
\begin{eqnarray}
\
e^+e^- \to H^+\,H^-\, h_i
\, ,\ \ \ 
\ e^+e^- \to h_i\, h_i\, A^0\,
, \ \ \ 
\ e^+e^- \to h^0\, H^0\, A^0\,,
\ \ \ \ \ \ \ \ \ (h_i=h^0,H^0,A^0) \label{3H}.
\end{eqnarray}
Let us recall that the general 2HDM\,\cite{hunter} is obtained by
canonically extending the SM Higgs sector with a second $SU(2)_L$
doublet with weak hypercharge $Y=1$, so that it contains $4$
complex scalar fields.
The free parameters $\lambda_i$ in the general, CP-conserving,
2HDM potential can be finally expressed in terms of the masses of
the physical Higgs particles ($M_{h^0}, M_{H^0}, M_{A^0},
M_{H^\pm}$), $\tan \beta$ (the ratio of the two VEV's $\langle
H_i^0\rangle$ giving masses to the up- and down-like quarks) and
the mixing angle $\alpha$ between the two $CP$-even states. There
remains, however, the coupling $\lambda_5$, which cannot be
absorbed in the previous quantities.  Following \,\cite{santi}, we
set $\lambda_5 = \lambda_6=2\sqrt{2}\,G_F\,M_{A^0}^2$. This
condition allows to keep closer to the MSSM structure of the Higgs
sector. Therefore we end up with $6$ free parameters, to wit:
$(M_{h^0},M_{H^0},M_{A^0},M_{H^{\pm}},\tan\alpha,\tan\beta)\,$.
Furthermore, to ensure the absence of tree-level flavor changing
neutral currents (FCNC), two main 2HDM scenarios arise: 1) type I
2HDM, in which one Higgs doublet couples only to down-like quarks,
whereas the other doublet does not couple to any quark; 2) type II
2HDM, where one doublet couples only to down-like  quarks and the
other doublet to up-like quarks. The MSSM Higgs sector is actually
a type II one, but of a very restricted sort (enforced by SUSY
invariance)\,\cite{hunter}.

Further constraints must be imposed to attest that the SM behavior
is sufficiently well reproduced up to the energies explored so
far, namely: $i)$ the perturbativity and unitarity bounds; $ii)$
the approximate $SU(2)$ custodial symmetry, i.e.
$|\delta\rho_{2HDM}|\le 10^{-3}$ \cite{custodial}, and $iii)$
consistency with the low-energy radiative $B$-meson decay (which
entails $M_{H^{\pm}}
> 350$ $GeV$ for $\tan \beta \ge 1$
\cite{gamba}, unless we consider type-I 2HDM). We refer the reader
to \cite{Ferrera:2007sp} for further details, in particular for
the full list of trilinear couplings within the general 2HDM that
are relevant for the present calculation.


\section{Double and triple Higgs boson production in the 2HDM} \label{sect:numerical}

In what follows we shall mainly discuss the leading-order 
3H production at the ILC within the general 2HDM
 and compare briefly
with the 2H production processes. Throughout the present work, we
have made used of the standard computational packages
\cite{feynarts}.

We begin by reporting on the tree-level results for the 2H
production within the 2HDM, specifically on the CP-conserving
channels $e^+e^- \to h^0\,A^0; H^0\,A^0; H^+H^-$. Let us firstly
define two different scenarios: \textbf{(a)} light, and
\textbf{(b)} heavy, Higgs boson masses (see Table \ref{tab:s1}).
We wish to use set I and III for the study of 2H production, while
set II and set III for the study of 3H production.

\begin{table}[h]
\footnotesize
\begin{center}
\begin{tabular}{|c||c|c|c|} \hline
\quad & Set I & Set II & Set III \\ \hline \hline
$M_{h^0}$\,$(GeV)$ & 100 & 100 & 200\\ \hline
$M_{H^{\pm}}$\,$(GeV)$ & 120 & 120 & 350\\\hline
$M_{H^0}$ \,$(GeV)$ & 150 & 150 & 250\\\hline
$M_{A^0}$\,$(GeV)$ & 140 & 300 & 340\\ \hline
\end{tabular}
\end{center}
\vspace*{-.2cm}
\caption{
Sets I, II and III of \textit{light} and
\textit{heavy} Higgs boson masses in the 2HDM. Sets I and III are
used for 2H production
and Sets II and III
for 3H production.
} \label{tab:s1}
\end{table}
%
We have found\,\cite{Ferrera:2007sp} that, in the light Higgs
boson mass regime, the 2H production rates within the 2HDM are
substantial (a few thousands events per $100$ fb$^{-1}$ of
integrated luminosity), with maximum values reaching
$\sigma(e^+\,e^- \to H^+\,H^-) \sim 0.1$ pb, whereas for heavier
Higgs masses (Set III) the optimal production rates lie around one
order of magnitude below. Nevertheless, even in these less favored
scenarios, the predicted rates are still quite sizeable within the
clean ILC environment. Worth noting is that the predicted 
cross-sections for the same 2H processes within the MSSM yield
$\sigma\sim 10^{-2}\,$pb and are thus comparable to the 2HDM
values for similar masses. This is a reflex of the fact that
formally the Higgs-Higgs-gauge boson couplings do not differ from
one model to the other. Therefore, we conclude that sizeable rates
of non-standard 2H production can be achieved at the ILC for both
SUSY and non-SUSY extended Higgs sectors, which implies that both
scenarios are hard to distinguish at the tree-level. A clear
separation of them  can only be accomplished through
the detailed study of radiative corrections to 2H production in
both the MSSM\,\cite{mssmloop} and the 2HDM\,\cite{ghk}.

Let us now discuss the case of the triple Higgs boson production
(cf. Eq.~(\ref{3H})) within the general 2HDM (see
Ref.~\cite{Ferrera:2007sp} for further details and discussions).
The basic result here is very different from the 2H case sketched
above, in the sense that the 3H cross-sections for the general
2HDM may carry large enhancements at the tree-level which are not
present in the MSSM. It means that the 3H channels in the general
2HDM could be, in contrast to the 2H ones, truly distinctive
already at the leading-order. The dynamical reason for this stems
from the structure of the trilinear Higgs boson couplings $HHH$ in
the general 2HDM. In contrast to the Yukawa coupling with
fermions, these trilinear couplings do not depend on whether we
are in type I or type II models, and can be largely enhanced in
certain regions of the parameter space (see Table 1 of
\cite{Ferrera:2007sp}). Such enhancements are not possible in the
MSSM case, owing to the purely gauge nature of these couplings 
in the SUSY case.

\begin{figure}[thb]
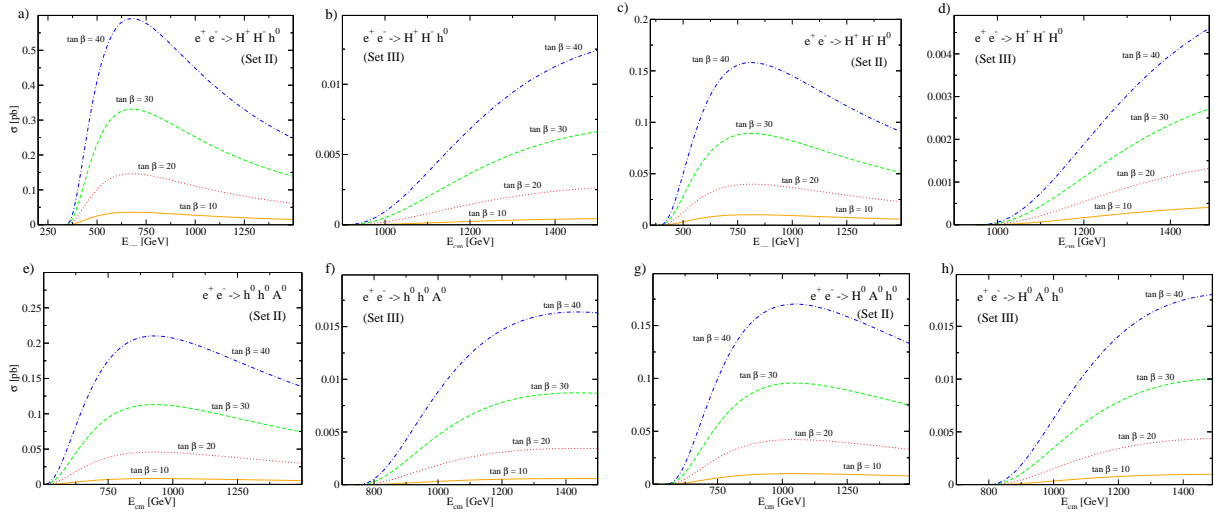

\centerline{
\begin{tabular}{llll}
$\!\!\!\!\!\!\!$\resizebox{!}{3.3cm}{\includegraphics{eeHHh_light.eps}} &
$\!\!\!\!\!\!\!\!\!\!\!$
\resizebox{!}{3.3cm}{\includegraphics{eeHHh_heavy.eps}} &
$\!\!\!\!\!\!\!\!\!\!$
%
\resizebox{!}{3.3cm}{\includegraphics{eeHHH0_light.eps}} &
$\!\!\!\!\!\!\!\!\!\!\!$
\resizebox{!}{3.3cm}{\includegraphics{eeHHH0_heavy.eps}} \\
$\!\!\!\!\!\!\!$\resizebox{!}{3.3cm}{\includegraphics{eehhA_light.eps}} &
$\!\!\!\!\!\!\!\!\!\!\!$
\resizebox{!}{3.3cm}{\includegraphics{eehhA_heavy.eps}} &
$\!\!\!\!\!\!$
\resizebox{!}{3.3cm}{\includegraphics{eeHAh_light.eps}} &
$\!\!\!\!\!\!\!\!\!\!\!$
\resizebox{!}{3.3cm}{\includegraphics{eeHAh_heavy.eps}}
\end{tabular}}
\vspace*{-.2cm}
\caption{\footnotesize{Total cross-section $\sigma$(pb)
for the triple Higgs boson production
processes $e^+\,e^- \to H^+\,H^-\,h^0$, $e^+\,e^- \to H^+\,H^-\,H^0$,
$e^+\,e^- \to h^0\,h^0\,A^0$ and $e^+\,e^- \to H^0\,A^0\,h^0$
in the general 2HDM as a
function of $\sqrt{s}$ and for different values of $\tan \beta$. In
each case the label of the process and the choice (Set II or Set
III) of Higgs boson masses used for the calculation is indicated,
see Table~{\protect\ref{tab:s1}}.
}} \label{fig:3higgs1}
\vspace*{-.3cm}
\end{figure}

The numerical analysis fully corroborates our expectations. In
Fig.\,\ref{fig:3higgs1} we have plotted the 3H production 
cross-sections within the 2HDM. Again, two different mass regimes (sets
II and III in Table \ref{tab:s1}) are considered. This is so
because of the constraints imposed by the radiative $B$-meson
decays on the charged Higgs boson masses for type II models
($M_{H^{\pm}}\gtrsim 350~\rm{GeV}$)\,\cite{gamba}. Lighter Higgs
boson regimes (allowed for type I models) entail optimal 
cross-sections at the level of $\sim 0.1$ pb or more, therefore implying
promising rates of at least $10^4$ events per $100~\rm{fb}^{-1}$
of integrated luminosity. The corresponding results for the heavy
Higgs boson scenario lie around $1-2$ orders of magnitude below
those for the light Higgs boson scenario, and they are attained at
higher values of $\sqrt{s}$. The results for set III (suited for
type II models) translate into rates of ${O}(10^2-10^3)$ events
per $100~\rm{fb}^{-1}$ of integrated luminosity, which may still
allow some comfortable detection of the signal in the clean ILC
environment. The remaining 3H channels (among the $7$
CP-conserving triple final states in (\ref{3H})) provide smaller
production rates in optimal conditions\,\cite{Ferrera:2007sp}.

In order to compare the 2HDM results with the corresponding MSSM
values, we have systematically searched for the (allowed) regions
across the MSSM parameter space where the optimal values for the
cross-sections are attained. Besides the exception triggered by
the resonant process  $e^+\,e^- \to h^0\,h^0\,A^0$, whose maximum
cross-section reads $\sigma(\sqrt{s}=1~TeV)\sim10^{-3}$ pb, the
remaining $\sigma(3H)$ are very small, namely they may reach
$\sigma(\sqrt{s}=1~TeV)\sim 10^{-6}$ pb at most. We can thus
assert that most of the 3H cross-sections in the MSSM are really
tiny and, hence, very difficult to detect in practice. Let us
finally stress that the extremely clean environment of the ILC
should allow a relatively comfortable tagging of the three Higgs
boson events for the typical 3H cross-sections that we have
obtained in the general 2HDM case.

\section{Discussion and conclusions}
\label{sect:conclusions}

We have computed the leading-order cross-sections for triple Higgs
boson production in a $e^+e^-$ linear collider -- cf.
Eq.~(\ref{3H}) -- within the general 2HDM. We have shown that the
production cross-sections may comfortably reach $0.1$pb within
type I 2HDM. Such optimal rates are achieved for sufficiently
large (or small) values of $\tan\beta$ ($\tan\beta$ $\gtrsim 20\,
,\, \tan\beta<0.1$). Moreover, in certain regions of the parameter
space, and for the most favorable processes (such as $e^+\,e^- \to
H^+\,H^-\,h^0$) the cross-sections can be pushed up to
$\sim\,1\,$pb. In spite of the fact that 3H production does not
involve any kind of Higgs-fermion interaction, and hence the
predicted rates should not depend on whether type I or type II
2HDM is considered, low-energy $B$-meson physics puts stringent
constraints on $M_{H^\pm}$ in the type II case. Such scenarios
with relatively heavy Higgs boson masses render maximum 
cross-sections that are roughly $10$ times smaller, i.e. of order of
$0.01\,$pb. The smaller number of events falls nonetheless in the
range of $10^3$ per $100\,$fb$^{-1}$ of integrated luminosity.
Remarkably, for both type I and type II 2HDM models the maximum
cross-sections lie far above their MSSM counterparts (which
typically remain at the tiny level of $\sigma\,\sim\,10^{-6}\,$pb).

Worth stressing is also the fact that, for the general 2HDM, the
maximum 3H production rates are fully comparable, and even larger
(for type I models), than the optimal values achieved by the 2H
processes ($\sigma\sim 0.1\,$pb). Such a result can be traced back
to the potential enhancement of the Higgs boson self-interactions
($HHH$) within the 2HDM, which cannot be realized in the MSSM case
because of the SUSY invariance of the interactions. We conclude
that the 3H final states in the general 2HDM can be a competitive,
if not the dominant, Higgs boson production mechanism at the ILC.
Owing to the extremely clean ILC environment, we expect that the
leading 3H signatures could hardly be  missed. The latter could be
quite spectacular since they should reveal in the form of $6$
heavy-quark jet final states. If a few, well identified, events of
this kind would be detected, it should hint strongly at
(non-SUSY) Higgs boson physics beyond the SM

\paragraph{Acknowledgments} GF thanks an ESR position of
the European network RTN MRTN-CT-2006-035505 Heptools, and the
hospitality of the HEP group at the Dept. ECM of the Univ. de
Barcelona. DLV has been supported by the MEC FPU grant Ref.
AP2006-00357; JG and JS in part by MEC and FEDER under project
FPA2007-66665 and also by DURSI Generalitat de Catalunya under
project 2005SGR00564. JG is thankful to the Universidad de
Zaragoza, for their kind hospitality.

\newcommand{\JHEP}[3]{ {JHEP} {#1} (#2)  {#3}}
\newcommand{\NPB}[3]{{\sl Nucl. Phys. } {\bf B#1} (#2)  {#3}}
\newcommand{\NPPS}[3]{{\sl Nucl. Phys. Proc. Supp. } {\bf #1} (#2)  {#3}}
\newcommand{\PRD}[3]{{\sl Phys. Rev. } {\bf D#1} (#2)   {#3}}
\newcommand{\PLB}[3]{{\sl Phys. Lett. } {\bf B#1} (#2)  {#3}}
\newcommand{\EPJ}[3]{{\sl Eur. Phys. J } {\bf C#1} (#2)  {#3}}
\newcommand{\PR}[3]{{\sl Phys. Rep } {\bf #1} (#2)  {#3}}
\newcommand{\RMP}[3]{{\sl Rev. Mod. Phys. } {\bf #1} (#2)  {#3}}
\newcommand{\IJMP}[3]{{\sl Int. J. of Mod. Phys. } {\bf #1} (#2)  {#3}}
\newcommand{\PRL}[3]{{\sl Phys. Rev. Lett. } {\bf #1} (#2) {#3}}
\newcommand{\ZFP}[3]{{\sl Zeitsch. f. Physik } {\bf C#1} (#2)  {#3}}
\newcommand{\MPLA}[3]{{\sl Mod. Phys. Lett. } {\bf A#1} (#2) {#3}}


\begin{thebibliography}{99}
%
\bibitem{Djouadi:1992pu} A. Djouadi, H.E. Haber and P.M. Zerwas,
\ZFP{57}{1993}{569};
%
A. Djouadi, H.E. Haber and P.M. Zerwas, \PLB{375}{1996}{203} ; A.
Djouadi, V. Driesen, W. Hollik and J. Rosiek, \NPB{491}{1997}{68};
J.L. Feng and T. Moroi, \PRD{56}{1997}{5962}.
%
\bibitem{mssmloop} V. Driesen, W. Hollik and J. Rosiek, (1996); S.
Heinemeyer, \IJMP{21}{2006}{2659}; A. Djouadi, W. Kilian, M.
Muhlleitner and P.M. Zerwas, \EPJ{10}{1999}{45}; T. Binoth, S.
Karg, N. Kauer and R. Ruckl, \PRD{74}{2006}{113008}.
%
\bibitem{ghk} J. Guasch, W. Hollik, A. Kraft, \NPB {B596} {2001}
{66}; J. Guasch, D. L\'opez-Val and J. Sol\`a, work in progress.
%
\bibitem{tripleMSSM}  P. Osland and P.N. Pandita, \PRD {59} {1998}{055013};
 A. Djouadi, W. Kilian, M. Muhlleitner and P.M. Zerwas,
\EPJ{10}{1999}{27}; D.J. Miller and S. Moretti, \EPJ {13} {2000}
{459}; F. Boudjema and A. Semenov, \PRD {66} {2002}{095007}.
%
\bibitem{Ferrera:2007sp}
  G.~Ferrera, J.~Guasch, D.~L\'opez-Val and J.~Sol\`a,
  \PLB{659}{2008}{297}.
%
\bibitem{hunter}J.F. Gunion, H.E. Haber, G.L. Kane and S. Dawson,
\textit{The Higgs hunter's guide}, Addison-Wesley, Menlo-Park, 1990.
%
\bibitem{santi} S. B\'ejar, J. Guasch and J. Sol\`a,
\NPB{600}{2001}{21}; \NPB{675}{2003}{270};
[\texttt{hep-ph/0101294}]; S. B\'ejar, PhD Thesis, UAB,
[\texttt{hep-ph/0606138}].
%
\bibitem{custodial} M.B. Einhorn, D. R. T. Jones and M. J. G. Veltman,
\NPB{123}{1977}{89}.
%
\bibitem{gamba} P. Gambino and M. Misiak, \NPB{611}{2001}{338};
M. Ciuchini, G. Degrassi, P. Gambino, and G. F. Giudice, \NPB{527}{1998}{21};
F. M. Borzumati and C. Greub, \PRD{58}{1998}{074004}.
%
\bibitem{feynarts} T. Hahn, \textit{FeynArts 3.2, FormCalc}
and \textit{LoopTools} user's guides, available from
\texttt{http://www.feynarts.de}; T. Hahn, \textit{Comput. Phys.
Commun.} \textbf{168} (2005) 78.


\end{thebibliography}
\end{document}